\documentclass[prb,a4paper,showpacs,preprintnumbers,showkeys,amsmath,amssymb]{revtex4}
\usepackage{graphicx}% Include figure files
\usepackage[T1]{fontenc}
\usepackage{dcolumn}% Align table columns on decimal point
\usepackage{bm}
\usepackage{natbib}
\begin{document}

\title{Influence of charge ordering on the magnetocaloric effect of (Pr,Ca)MnO$_3$ manganites}

\author{M.S. Reis}
\email{marior@fis.ua.pt}
\author{V.S. Amaral}
\affiliation{Departamento de Física and CICECO, Universidade de
Aveiro, 3810-193 Aveiro, Portugal}
\author{J.P. Araújo}
\affiliation{IFIMUP, Departamento de Física, Universidade do
Porto, 4150 Porto, Portugal}
\author{P.B. Tavares}
\affiliation{Departamento de Química, Universidade de
Trás-os-Montes e Alto Douro, 5001-911 Vila Real, Portugal}
\author{A.M. Gomes and I.S. Oliveira}
\affiliation{Centro Brasileiro de Pesquisas Físicas, Rua Dr.
Xavier Sigaud 150 Urca, 22290-180 Rio de Janeiro-RJ, Brasil
}%

\date{\today}

\preprint{Submitted to Phys. Rev. B.}
\begin{abstract}
In the present work we analyze the influence of the
charge-ordering on the magnetocaloric effect of
Pr$_{1-x}$Ca$_x$MnO$_3$ manganites. The results for the samples
with $x<$0.30 present the usual ferromagnetic behavior, peaking at
the Curie temperature T$_C$. In contrast, for the samples above
the onset concentration for the charge-ordering ($x\sim$0.30), an
anomalous magnetic entropy change was observed below the
charge-ordering temperature $T_{CO}$, persisting for lower
temperatures. This effect is associated to a positive contribution
to the magnetic entropy change due to charge-ordering, which is
superimposed to the negative contribution from the spin ordering.
We found that the charge-ordering component peaks at $T_N$ and is
negligible above $T_{CO}$. Moreover, around T$^\star$ (temperature
below which the insulator-metal transition induced by magnetic
field becomes completely irreversible, where the system remains in
such state even after the external magnetic field has been
removed), we found extremely large values for the magnetic entropy
change, establishing a `colossal' magnetocaloric effect on CMR
manganites.

\end{abstract}

\pacs{75.30.Sg, 75.47.Lx, 75.47.Gk} \keywords{Manganites,
magnetocaloric effect, charge ordering}

\maketitle
\section{Introduction}\label{section_introducao}

The magnetocaloric effect (MCE) is intrinsic to magnetic
materials, and is induced via coupling of the magnetic sublattice
with the magnetic field, which alters the magnetic part of the
total entropy due to a corresponding change of the magnetic field.
The MCE can be estimated via the magnetic entropy change $\Delta
S_{\textnormal{\tiny{M}}}(T,\Delta H)$, and is a function of both,
temperature $T$ and the magnetic field change $\Delta H$, being
usually recorded as a function of temperature, at a constant
$\Delta H$. In addition, the MCE has a significant technological
importance, since magnetic materials with large MCE values could
be employed in various thermal devices\cite{JAP_86_1999_565}.

The magnetic entropy is related to the magnetization $M$, magnetic
field strength $H$ and absolute temperature $T$ through the
Maxwell relation \cite{livro_Morrish}:
\begin{equation}\label{maxwell_relatation}
  \left(\frac{\partial S_{\textnormal{\tiny{M}}}(T,H)}{\partial
  H}\right)_T = \left(\frac{\partial M(T,H)}{\partial
  T}\right)_H
\end{equation}
which after integration yields:
\begin{equation}\label{DS1}
  \Delta S_{\textnormal{\tiny{M}}}(T,\Delta H) =
  \int\limits^{H_F}_{H_I} dS_{\textnormal{\tiny{M}}}(T,H)_T =
  \int\limits^{H_F}_{H_I} \left(\frac{\partial M(T,H)}{\partial T}\right)_H dH
\end{equation}

Hence, $\Delta S_{\textnormal{\tiny{M}}}(T,\Delta H)$ can be
numerically calculated for any temperature, using Eq.\ref{DS1} and
the measured magnetization as a function of magnetic field and
temperature. Generally, since temperature stabilization is the
longest step in the process of collecting magnetization data, the
measurements are usually carried out isothermally by varying the
magnetic field, for a range of temperature values.

Several authors, through many decades, have studied the
magnetocaloric effect in a large variety of magnetic materials.
However, more recently, an enormous amount of work
\cite{{JMMM_238_2002_25},{JMMM_219_2000_183},
{JAP_90_2001_5689},{JAP_84_1998_3798},{APL_69_1996_3596},
{JMMM_242-245_2002_668},{JMMM_242_2002_698},{EL_52_2000_589},{JAP_91_2002_8903},
{JAP_91_2002_9943},
{JAP_79_1996_373},{APL_77_2000_1026},{PRL_78_1997_1142},
{JMMM_173_1997_302},{APL_73_1998_390},{JMMM_208_2000_85}} were
devoted to explore the MCE in the mixed-valency manganites
AMnO$_3$, where $A$ is a trivalent rare-earth mixed with a
divalent alkaline-earth.

Particularly interesting as candidates to technological
applications are the Pr$_{1-x}$Ca$_x$MnO$_3$ manganites, since
their phase diagram exhibit a rich variety of magnetic, electric
and crystallographic structures. In this direction, we aim to
explore the magnetocaloric effect through the concentration range
of phase competition between the
antiferromagnetic-insulator-charge-ordered state and
ferromagnetic-insulator state, and analyze the influence of the
charge ordering on the magnetocaloric properties of these
compounds.

Hence, the next section is devoted to a brief survey of the
magnetic, electric and crystallographic properties of the
Pr$_{1-x}$Ca$_x$MnO$_3$ manganites. Following, the section after
is dedicated to the experimental details. In section
\ref{section_results} a complete description of the experimental
results is given, with a thorough discussion and analysis in the
following sections.

\section{A brief survey}\label{section_brief_survey}

For $x<$0.30 an orthorhombic $O'$ (c/$\sqrt{2}\lesssim$a$<$b)
crystal structure establishes below 950 K, for $x=$0.0, and 325 K,
for $x\sim$0.30, whereas for 0.30$<x<$0.75 the crystal phase
arisen is a pseudo-tetragonal compressed $T$ (c/$\sqrt{2}<$a), for
temperatures below the onset temperature for the charge-ordering
T$_{CO}$. For high temperatures, the crystal phase is always
orthorhombic of $O$ type
(a$\simeq$b$\simeq$c/$\sqrt{2}$)\cite{JMMM_53_1985_153}. For
$x\sim$0.30, at low temperatures, a strong crystal phase
competition arises between the orthorhombic $O'$ and the
pseudo-tetragonal compressed $T$ \cite{JMMM_53_1985_153}.

However, the phase coexistence around $x\sim$0.30 is not limited
to the crystallographic aspects, but a remarkable electric
\cite{NMR_prca} and magnetic
\cite{JAP_79_1996_5288,{PRB_57_1998_3305},{PRB_62_2000_3381},
{PRB_62_2000_13876},{PRB_63_2001_224403},{PRB_53_1996_r1689},
{PRB_60_1999_12191},{MSEB_63_1999_22},{PMB_81_2001_417}} phase
competition can also be found. In this direction, neutron
diffraction \cite{PRB_57_1998_3305} and muon spin relaxation
($\mu$SR) \cite{PRB_62_2000_3381} recognized, for $x\sim$0.30, two
magnetic transition, around 140 K and 120 K. The first transition,
at higher temperature, corresponds to the antiferromagnetic
arrangement, and the other represents an ferromagnetic
contribution. Hence, two different interpretations are, in
principal, possible: two separated magnetic phases with different
critical temperatures, or, alternatively, a
collinear-antiferromagnetic phase with T$_N$ at higher
temperature, with an additional magnetic transition to a
canted-antiferromagnetic structure, at lower temperature. However,
de Gennes \cite{PR_118_1960_141} stressed out that an uniform
canting can be achieved only in the presence of free carriers,
i.e., in a metal, whereas the present system is an insulator.
Bound carriers would give rise to a local ferromagnetic distortion
of the spin system, which would therefore become inhomogeneous
\cite{PR_118_1960_141}. In an insulator, the bound electrons would
form small ferromagnetic cluster in a antiferromagnetic matrix,
and the cluster of spins would then align parallel to each other
at T$_C$ \cite{PRB_57_1998_3305}. Additionally, the presence of
the phase boundary between two different crystallographic types
($O'$ and $T$), lead us to conclude that the coexistence of two
phases is more probable than a canted-antiferromagnetic structure.
Finally, NMR \cite{NMR_prca}, $\mu$SR \cite{PRB_62_2000_3381} and
neutron diffraction \cite{PRB_57_1998_3305,{JMMM_53_1985_153}},
add and support our suggestion. Thus, this view will be assumed in
the further discussions.

For $x<$0.15, a spin-canted insulator CI phase establishes below
100 K
\cite{PRB_53_1996_r1689,{JAP_79_1996_5288},{MSEB_63_1999_22},{JMMM_53_1985_153}},
whereas for 0.15$<x<$0.30 a ferromagnetic insulator FMI phase
arises, with Curie temperature around 120 K
\cite{PRB_60_1999_12191,{PRB_53_1996_r1689},{JAP_79_1996_5288},{MSEB_63_1999_22}}.
For 0.30$<x<$0.80, an antiferromagnetic-insulator AFMI phase
arises for temperatures typically below 170 K
\cite{JMMM_53_1985_153,{PRB_60_1999_12191},{PRB_53_1996_r1689}},
coexisting with a charge-ordered CO state with onset temperature
T$_{CO}$ between 210 K, for x=0.30, and 170 K, for x=0.80
\cite{PRB_60_1999_12191}. Additionally, it is well
established\cite{PRB_57_1998_3305,{PRB_62_2000_3381},{JMMM_53_1985_153}}
that the clusters embedded in the antiferromagnetic matrix achieve
the ferromagnetic order around 110 K, for $x=$0.30, and 42 K, for
$x=0.40$ \cite{PRB_53_1996_r1689}. In this direction, our recent
work using NMR \cite{NMR_prca}, gives the ferromagnetic fraction
within the antiferromagnetic matrix, as a function of Ca content,
$x$.

For temperatures below than a characteristic temperature
T$^\star$, ranging from 60 K for $x=$0.30 and 20 K for $x=$0.35
\cite{PRB_53_1996_r1689}, the application of a magnetic field
induces a first-order and completely \emph{irreversible}
transition to a fully ferromagnetic-metallic FMM state
\cite{JPSJ_64_1995_3626,{JAP_79_1996_5288},{PRB_53_1996_r1689},{JMMM_200_1999_1}},
where the system remains in such state even after the external
magnetic field has been removed. However, for temperature ranges
T$^\star<$T$<$T$_{CO}$, the system return to the insulator state,
with hysteresis, after the increase-decreasing magnetic field
cycle. The strong hysteresis associated with this transition is
indicated by the shaded region in the $H-T$ diagram sketched in
figure 1, for $x=$0.35. Additionally, this insulating state can
also be driven metallic by an applied electric field
\cite{nature_388_1997_50,{PRB_61_2000_11236}}, high pressure
\cite{PRB_55_1997_7549}, visible light
\cite{science_280_1998_12,{PRL_78_1997_4257}} or x-ray
\cite{nature_386_1997_813,{PRB_57_1998_3305}}.
\begin{figure}
\begin{center}
\includegraphics[width=10cm]{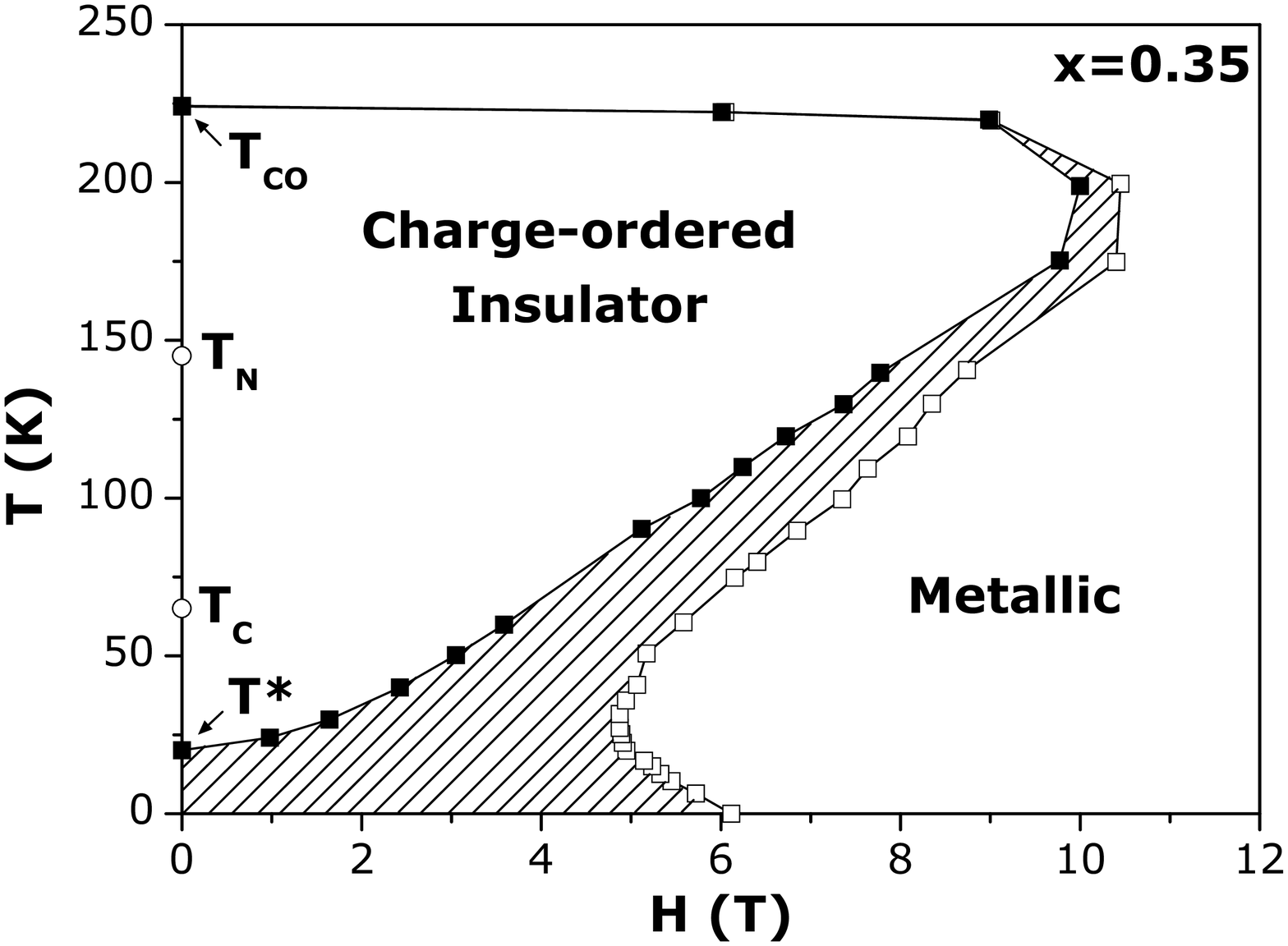}
\end{center}
\caption{$H-T$ phase diagram for Pr$_{1-x}$Ca$_x$MnO$3$, with
$x=$0.35. The insulator-metal transition induced by an external
magnetic field is reversible, with hysteresis, for temperatures
between T$_{CO}$ and T$^\star$, as indicated by the shaded area.
Below T$^\star$, the insulator state can also be driven metallic,
and kept in this state even after the magnetic field have been
removed. After Tomioka \emph{et al.} \cite{PRB_53_1996_r1689}.}
\label{figure1}
\end{figure}

However, all values mentioned above are slightly dependent on the
sample preparation procedure, and the differences in results can
be attributed to the grain size
\cite{JAP_91_2002_9943,{JMMM_221_2000_57},{PRB_62_2000_6437},{JMMM_189_1998_321}},
oxygen content
\cite{JMMM_189_1998_321,{PRB_54_1996_6172},{PRB_62_2000_11328},{JMMM_167_1997_200},{LTP_27_2001_283}},
vacancy on the lattice \cite{PRB_57_1998_1024}, etc. Additionally,
the phase diagram presented here (figure 2), is not completely
established, since the magnetic structure for several values of Ca
concentration $x$ is still a matter of discussion
\cite{JAP_79_1996_5288,{PRB_57_1998_3305},{PRB_62_2000_3381},
{PRB_62_2000_13876},{PRB_63_2001_224403},{PRB_53_1996_r1689},
{PRB_60_1999_12191},{MSEB_63_1999_22},{PMB_81_2001_417}}.
\begin{figure}
\begin{center}
\includegraphics[width=10cm]{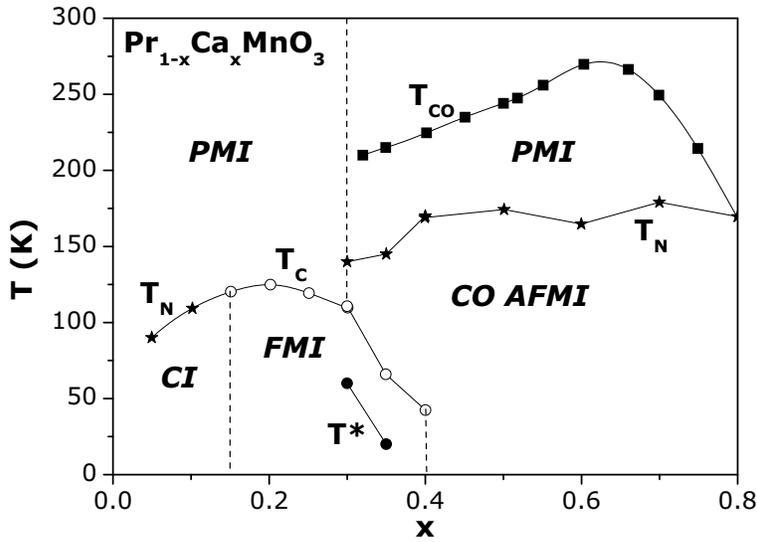}
\end{center}
\caption{Magnetic and electric phase diagram for
Pr$_{1-x}$Ca$_x$MnO$_3$ manganites. $PMI$ - paramagnetic
insulator; $CI$ - spin canted insulator; $FMI$ - ferromagnetic
insulator; $CO$ $AFMI$ - charge-ordered antiferromagnetic
insulator.} \label{figure2}
\end{figure}

\section{Experimental Procedure}\label{section_experimental_procedure}

The samples $x=$ 0.20, 0.25, 0.30, 0.32, 0.35 and 0.40, were
prepared by the ceramic route, starting from the stoichiometric
amount of Pr$_2$O$_3$ (99.9 \%), CaCO$_3$ (>99 \%) and MnO$_2$
(>99 \%), and heated in air, with five intermediate
crushing/pressing steps. The final crushed powders were compressed
and sintered in air at 1350 $^\textnormal{o}$C during 45 hours,
with a subsequent fast freezing of the samples. X-ray diffraction
patterns confirmed that the samples lie in the \textit{Pbnm} space
group, without vestige of spurious phase.

The temperature and external magnetic field dependence of the
magnetization were carried out using a commercial SQUID
magnetometer. The data were acquired after the sample had been
zero-field cooled, under the isothermal regime (M \emph{vs.} H
curves), varying the applied magnetic field from zero up to 50
kOe, and temperature ranges from 10 K up to 400 K. After each M
\emph{vs.} H curve, the sample was heated without the influence of
the external magnetic field. For the DC-susceptibility
$\chi_{\textnormal{\tiny{DC}}}$ ($=$M/H at low field), the
measurements were performed at a fixed magnetic field (H=10 Oe),
sweeping the temperature.

\section{Results}\label{section_results}

In this section we will describe the field and temperature
dependence of the magnetization M(T,H), since it is one of the
best experimental tools to understand the magnetocaloric potential
of the system under study.

The zero-field-cooled (ZFC) and field-cooled (FC)
DC-susceptibility $\chi_{\textnormal{\tiny{DC}}}$ (=M/H, with H=10
Oe), were measured for all samples available. We observed a
similar behavior for those below the onset concentration for
charge ordering ($x\sim$0.30), with a well defined transition from
the paramagnetic to the ferromagnetic phase. While the
DC-susceptibility $\chi_{\textnormal{\tiny{DC}}}$, after field
cooling, increase with decreasing temperature, after the sample
had been zero-field-cooled, it keeps an almost temperature
independence feature. Additionally, from the quantity
$1/\chi_{\textnormal{\tiny{DC}}}$, we could observe the usual
Curie-Weiss law, allowing to estimate the value of the
paramagnetic effective moment $p_{eff}$ and the paramagnetic Curie
temperature $\theta_p$. Thus, figure 3 sketched the behavior above
described, representative for all samples $x<$0.30.
\begin{figure}
\begin{center}
\includegraphics[width=10cm]{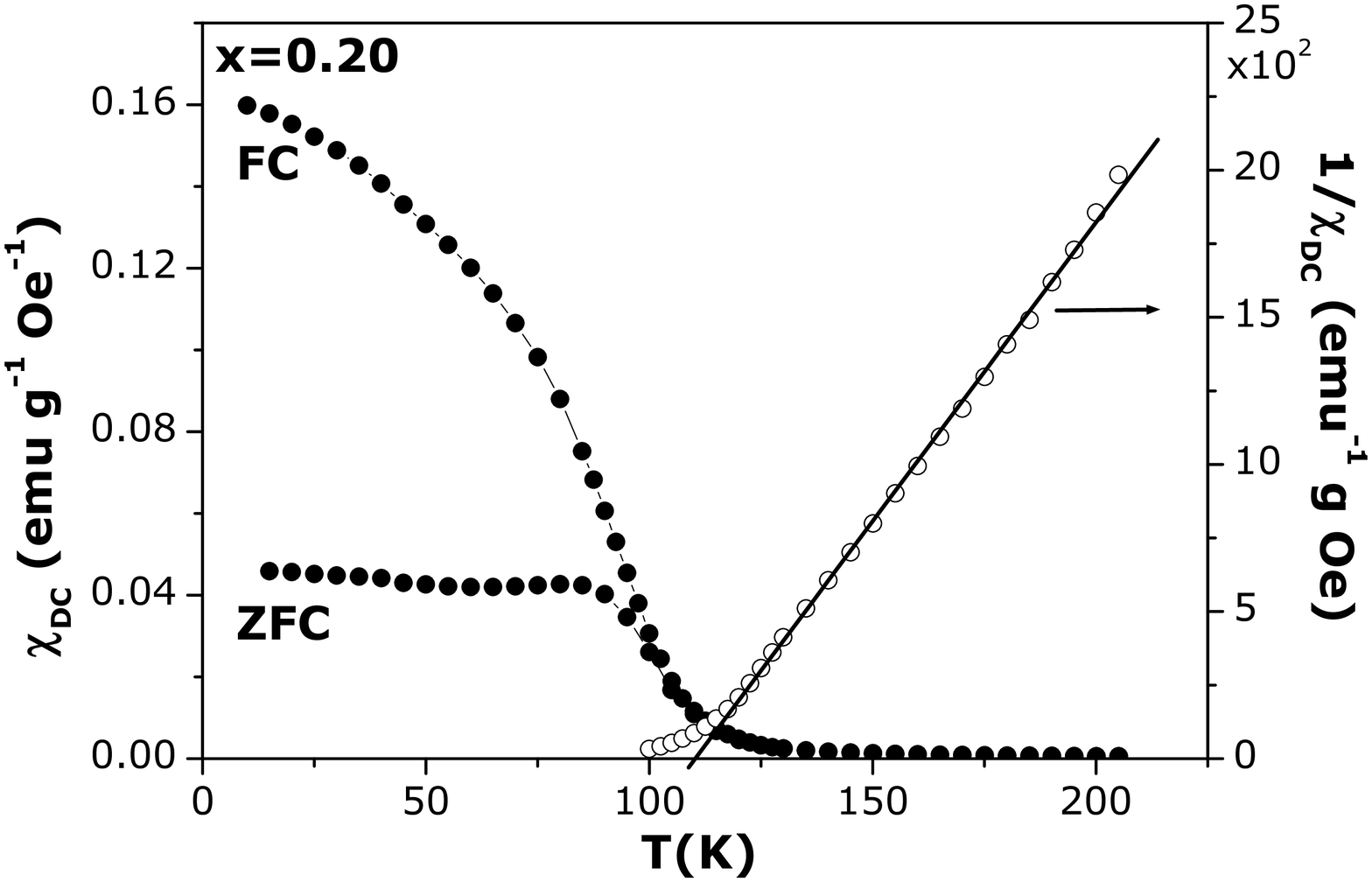}
\end{center}
\caption{Left axis: zero-field-cooled (ZFC) and field-cooled (FC)
DC-susceptibility $\chi_{\textnormal{\tiny{DC}}}$ (=M/H, with H=10
Oe) as a function of temperature, for $x=$0.20. Right axis:
temperature dependence of the inverse of the DC-susceptibility.}
\label{figure3}
\end{figure}

On other hand, for $x>$0.30 the DC-susceptibility change its slope
when the sample is cooled down through the CO transition, implying
in two different Curie-Weiss law, depending on the temperature
range: one for T$>$T$_{CO}$ and other for T$_N<$T$<$T$_{CO}$. This
behavior, already reported \cite{PRB_62_2000_3381}, is illustrated
in figure 4, for $x=$0.40. In addition, a well defined transition
from the paramagnetic phase to the antiferromagnetic phase are
also observed, as well as the characteristic temperature
T$^\star$, below which the field-induced insulator-metal
transition is completely irreversible (see section
\ref{section_brief_survey}). The quantities obtained from the
analysis of the DC-susceptibility are summarized on table I.
Figure 5 sketched the concentration dependence of (a) the
paramagnetic Curie temperature $\theta_p$ and (b) the paramagnetic
effective moment $p_{eff}$. The two branches curve displayed is
consequence of the two Curie-Weiss law found for $x>$0.30.
\begin{figure}
\begin{center}
\includegraphics[width=10cm]{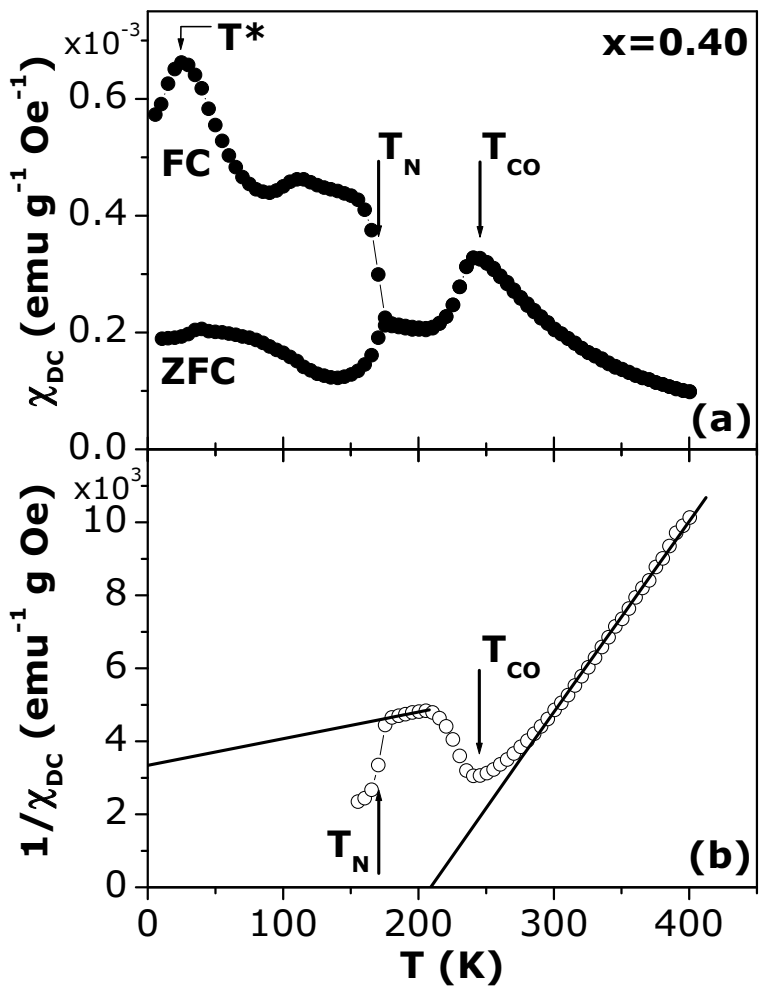}
\end{center}
\caption{ (a) Temperature dependence of the zero-field-cooled
(ZFC) and field-cooled (FC) DC-susceptibility
$\chi_{\textnormal{\tiny{DC}}}$ (=M/H, with H=10 Oe), for
$x=$0.40. (b) Temperature dependence of the inverse of the
DC-susceptibility, with two distinct Curie-Weiss law: one for
T$>$T$_{CO}$, and other for T$_N<$T$<$T$_{CO}$.} \label{figure4}
\end{figure}
\begin{figure}
\begin{center}
\includegraphics[width=10cm]{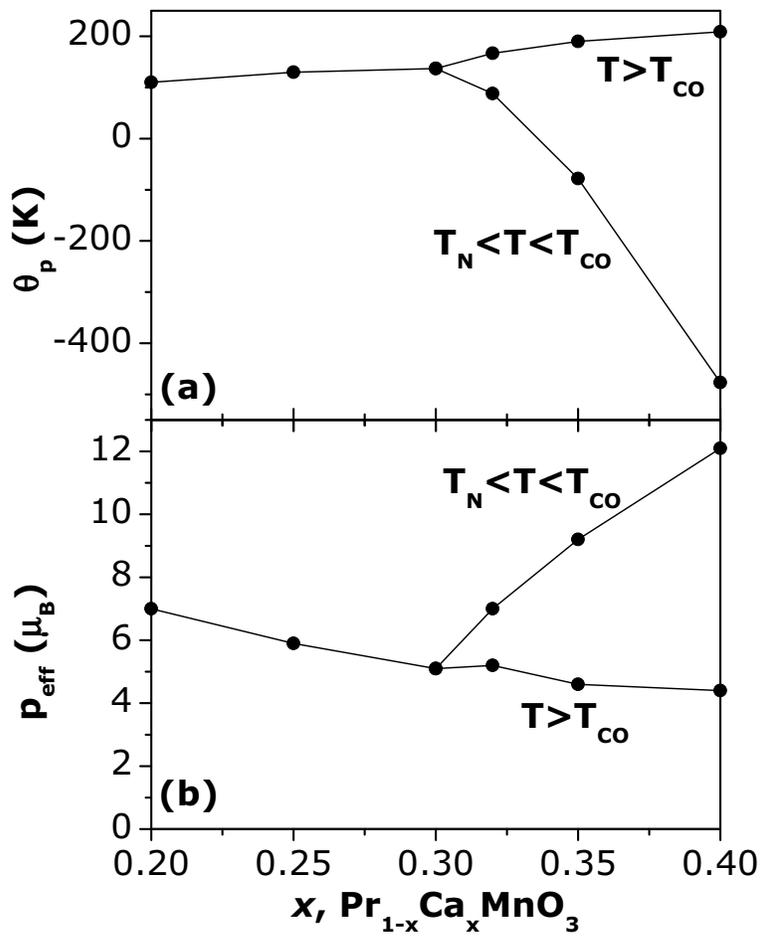}
\end{center}
\caption{(a) Paramagnetic Curie temperature $\theta_p$ and (b)
paramagnetic effective moment $p_{eff}$ as a function of Ca
content, $x$.} \label{figure5}
\end{figure}

\begin{table}[b]\label{table_tc}
\caption{Values obtained for T$^\star$, the onset temperature
below which the field-induced insulator-metal transition is
completely irreversible; T$_{CO}$, charge-ordering temperature;
the critical temperature T$_{crit}$, obtained form the maximum of
$d\chi_{\textnormal{\tiny{DC}}}/dT$; $\theta_p$, the paramagnetic
Curie temperature; and, finally, $p_{eff}$, the paramagnetic
effective moment.}
\begin{ruledtabular}
\begin{tabular}{cccccc}
 $x$ & T$^\star$ (K) & T$_{CO}$ (K) & T$_{crit}$ (K) & $\theta_p$ (K) & p$_{eff}$ ($\mu_B$)  \\
\hline
0.20 & - & - & 100\footnotemark[1] & 110 & 7.0 \\
0.25 & - & - & 120\footnotemark[1] & 130 & 5.9 \\
0.30 & - & - & 128\footnotemark[1] & 137 & 5.1 \\
0.32 & 26 & 210 & 113\footnotemark[2] &
\begin{tabular}{c}
88\footnotemark[3] \\
167\footnotemark[4] \\
\end{tabular}
&
\begin{tabular}{c}
7.0\footnotemark[3] \\
5.2\footnotemark[4] \\
\end{tabular}
\\
0.35 & 19 & 222 & 152\footnotemark[2] &
\begin{tabular}{c}
-78\footnotemark[3] \\
190\footnotemark[4] \\
\end{tabular}
&
\begin{tabular}{c}
9.2\footnotemark[3] \\
4.6\footnotemark[4] \\
\end{tabular}
\\
0.40 & 11 & 244 & 170\footnotemark[2] &
\begin{tabular}{c}
-477\footnotemark[3] \\
209\footnotemark[4] \\
\end{tabular}
&
\begin{tabular}{c}
12.1\footnotemark[3] \\
4.4\footnotemark[4] \\
\end{tabular}
\\
\end{tabular}
\end{ruledtabular}
\footnotetext[1]{Curie temperature} \footnotetext[2]{Néel
temperature} \footnotetext[3]{T$_N<$T$<$T$_{CO}$}
\footnotetext[4]{T$>$T$_{CO}$}
\end{table}

The temperature and field dependence of the magnetization M(T,H)
were also measured for all samples. From the data analysis of the
several M \emph{vs.} H isotherms, we could build the curves for
the thermal dependence of the magnetization, at a fixed magnetic
field. For $x=$0.20, as the temperature is further decreased, one
observes that the magnetization starts to increase faster below
100 K, peaking at around 35 K, as displayed in figure 6(a). It can
be related to the vicinity of this sample to the onset
concentration to spin-canted order ($x<$0.15)
\cite{JAP_79_1996_5288,{PRB_53_1996_r1689}}. On the other hand,
for $x=$0.25 and 0.30 an usual behavior for the thermal dependence
of the magnetization is found, as in figure 6(b).
\begin{figure}
\begin{center}
\includegraphics[width=10cm]{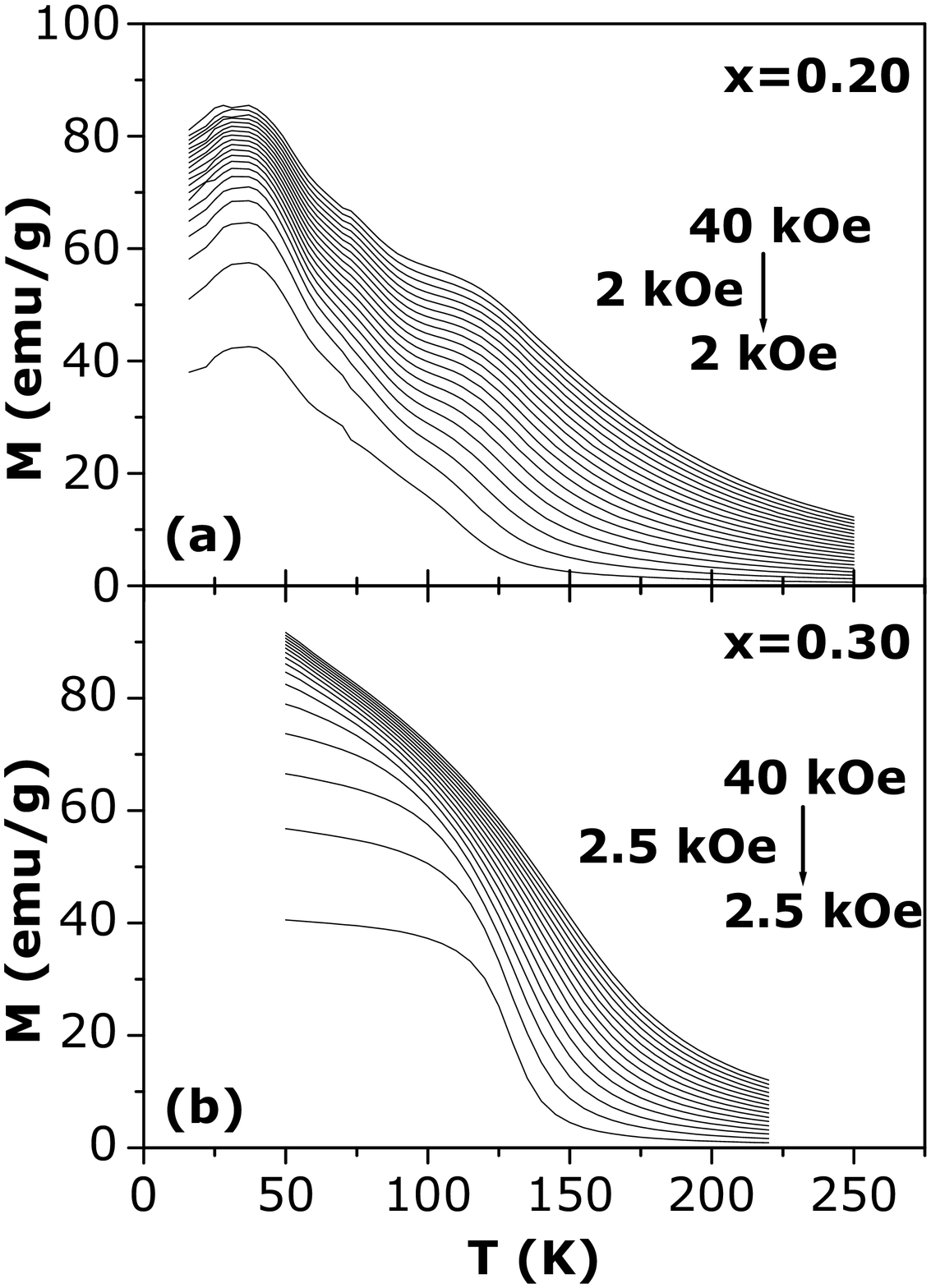}
\end{center}
\caption{Temperature dependence of the magnetization data, for
(a)$x=$0.20 and (b)$x=$0.30. The graphic was built from the
measured M \emph{vs.} H curves, for several temperatures.}
\label{figure6}
\end{figure}

However, for Ca concentration $x$ above the onset concentration
for charge ordering ($x\sim$0.30), the temperature dependence of
the magnetization have interesting features. For $x=$0.32, for
instance, the two peaks around 220 K and 130 K correspond to the
establishment of the charge ordering and the antiferromagnetic
spin ordering, respectively, in accordance with several works
\cite{JAP_79_1996_5288,{PRB_53_1996_r1689}}, including those using
neutron diffraction
\cite{JMMM_53_1985_153,{PRB_58_1998_8694},{PRB_52_1995_13145},{PRB_57_1998_3305}}.
As the temperature is further decreased, we can verify a
remarkable increasing of the magnetization below 50 K, peaking at
around T$^\star$=26 K, with a subsequent lost of magnetic moment,
reaching 40 emu/g at 5 K and 40 kOe. A similar behavior had been
found for all samples with $x>$0.30. This features, that can be
seen in figure 7 for (a) $x=$0.32 and (b) $x=$0.40, was already
observed for $x=$0.37 \cite{PRB_63_2001_224403}.
\begin{figure}
\begin{center}
\includegraphics[width=10cm]{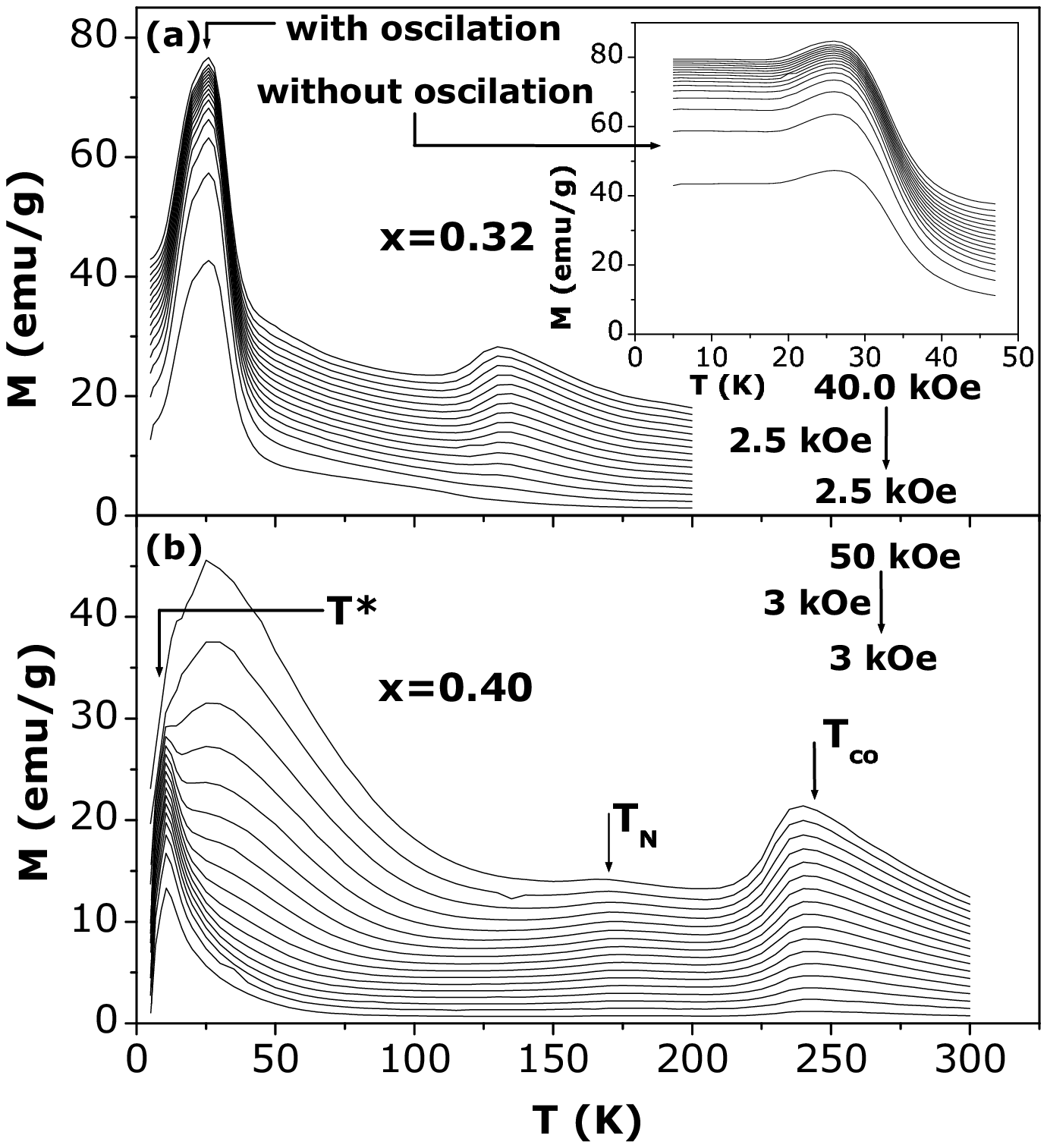}
\end{center}
\caption{Temperature dependence of the magnetization data, for
(a)$x=$0.32 and (b)$x=$0.40. The graphic was built from the
measured M \emph{vs.} H curves. For all measurements, when the
magnetic field is turned off, it oscillates around zero field in
order to avoid residual field to the next M \emph{vs.} H curve,
except for the data presented in the inset.} \label{figure7}
\end{figure}

For all measurements, when the magnetic field is turned off, it
oscillates around zero field in order to avoid residual field to
the next M \emph{vs.} H curve. If this procedure is not performed,
a remarkable difference is found for the magnetization values at
low temperatures. The M \emph{vs.} T curves also peaks around
T$^\star$=26 K, although keeps a high magnetization value for
temperature below the peak (80 emu/g for 5 K and 40 kOe). This
feature can be seen in the inset of the figure 7(a).

Following, the next section is devoted to analyze the influence of
the charge-ordering on the magnetocaloric effect, which will be
derived from the magnetization data, using Eq.\ref{DS1}.

\section{Influence of the Charge-Ordering on the Magnetocaloric Effect}\label{section_influence_charge-orderig_MCE}

In the previous sections, we provided detailed information
referred to the magnetic properties of the Pr$_{1-x}$Ca$_x$MnO$_3$
manganites. The aim of this section is the discussion concerning
the analysis of the magnetocaloric potential of these manganites,
obtained from the previously presented magnetic data.

\subsection{MCE around T$_{crit}$}

The magnetic entropy change $\Delta S_{\textnormal{\tiny{M}}}$(T)
can be estimated using Eq.\ref{DS1}, and are sketched in figure 8
for all samples available ($x=$0.20; 0.25; 0.30; 0.32; 0.35;
0.40).
\begin{figure}
\begin{center}
\includegraphics[width=10cm]{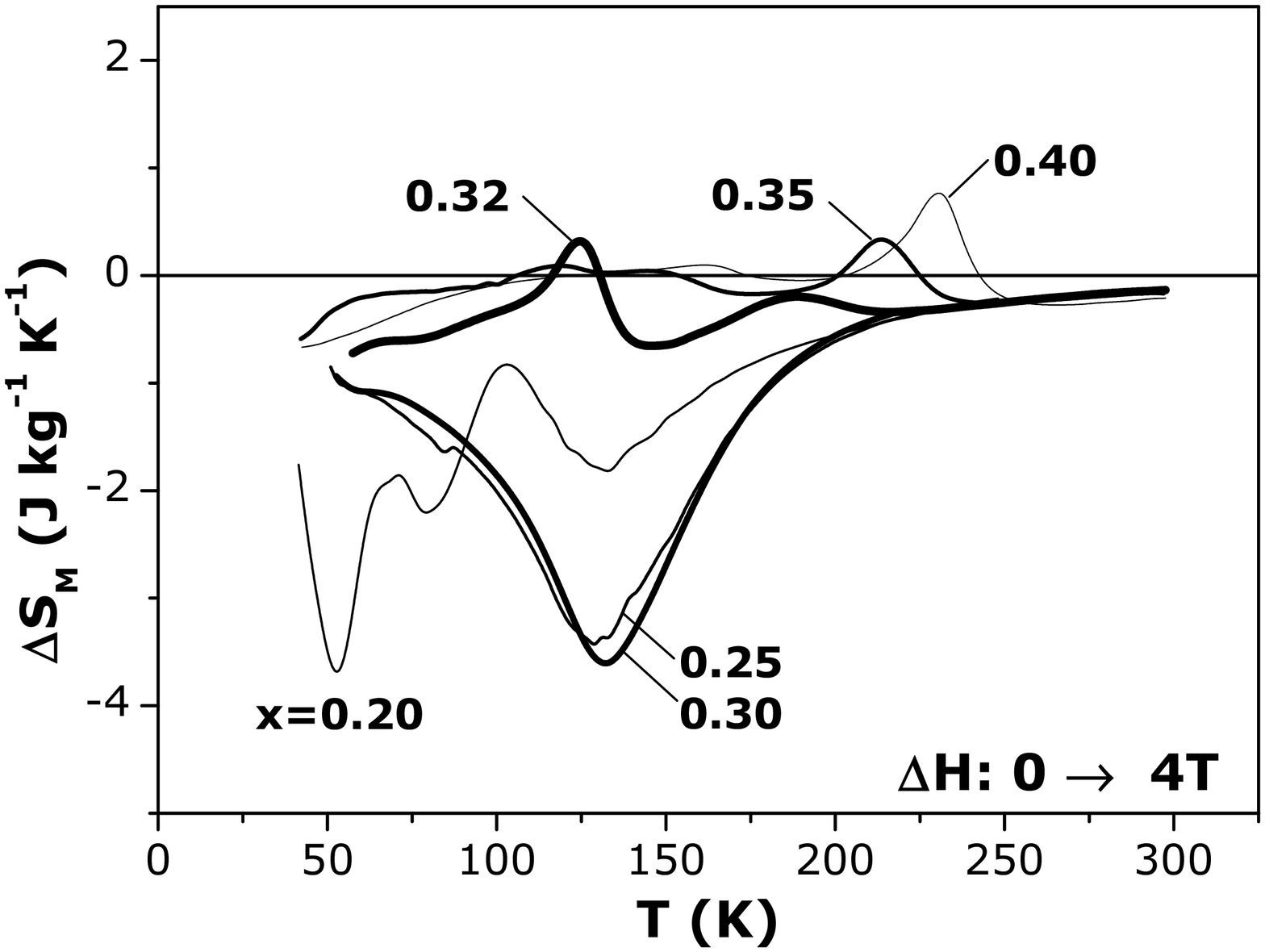}
\end{center}
\caption{Temperature dependence of the magnetic entropy change,
around T$_{crit}$, under 4 T of magnetic field change, for
$x=$0.20; 0.25; 0.30; 0.32; 0.35 and 0.40.} \label{figure8}
\end{figure}

For $x=$0.20, the magnetic entropy change has an usual behavior
for temperatures above 100 K, where the magnetization data (figure
6(a)) has also a well shaped feature. However, below 100 K, the
magnetization starts to increase faster as the temperature is
further decreased, as already discussed in Section IV, and it is
possibly related to the closed vicinity of this sample to the
spin-canted structure that arise below $x=$0.15
\cite{PRB_53_1996_r1689,{JAP_79_1996_5288},{MSEB_63_1999_22},{JMMM_53_1985_153}}
(see figure 2). Thus, the anomalous behavior found for T$<$100 K
on the magnetic entropy change of $x$=0.20 can also be explained
considering the influence of the spin-canted structure. On the
other hand, for $x=$0.25 and 0.30, concentrations completely
embedded within the ferromagnetic region, a very usual behavior
for $\Delta S_{\textnormal{\tiny{M}}}$ are found for both, as
sketched in figure 8. We consider, in the case of $x=$0.25 and
0.30, that the contribution to the magnetic entropy change is
purely due to the spin magnetic moment of the sample. However, for
samples with concentrations above $x\sim$0.30, the charge-ordering
arrangement plays a decisive role. When the temperature is further
decreased, the magnetic entropy change $\Delta
S_{\textnormal{\tiny{M}}}$ follows an usual shape until reach the
onset temperature for the charge-ordering $T_{CO}$, below which
such behavior is completely broken, as can be observed in figure
8.

To analyze this intriguingly feature, we consider two different
contributions to the total magnetic entropy change $\Delta
S_{\textnormal{\tiny{M}}}$: one refers to the spin rearrangement
$\Delta S_{\textnormal{\tiny{spin}}}$, and the other concerns to
the charge ordering rearrangement $\Delta
S_{\textnormal{\tiny{CO}}}$, as follow:

\begin{equation}\label{Ds_contribuitions}
    \Delta S_{\textnormal{\tiny{M}}}=\Delta S_{\textnormal{\tiny{spin}}}+\Delta S_{\textnormal{\tiny{CO}}}
\end{equation}

Thus, considering that the spin contribution to the total magnetic
entropy change of $x>$0.30 is almost similar to the purely spin
contribution of $x=$0.30, shifted to its T$_{crit}$, we can
satisfactorily estimate the charge-ordering contribution. Thus,
figure 9 presents the CO and spin contribution to the total
magnetic entropy change $\Delta S_{\textnormal{\tiny{M}}}$, for
(a)$x=$0.32, (b)0.35 and (c)0.40.
\begin{figure}
\begin{center}
\includegraphics[width=10cm]{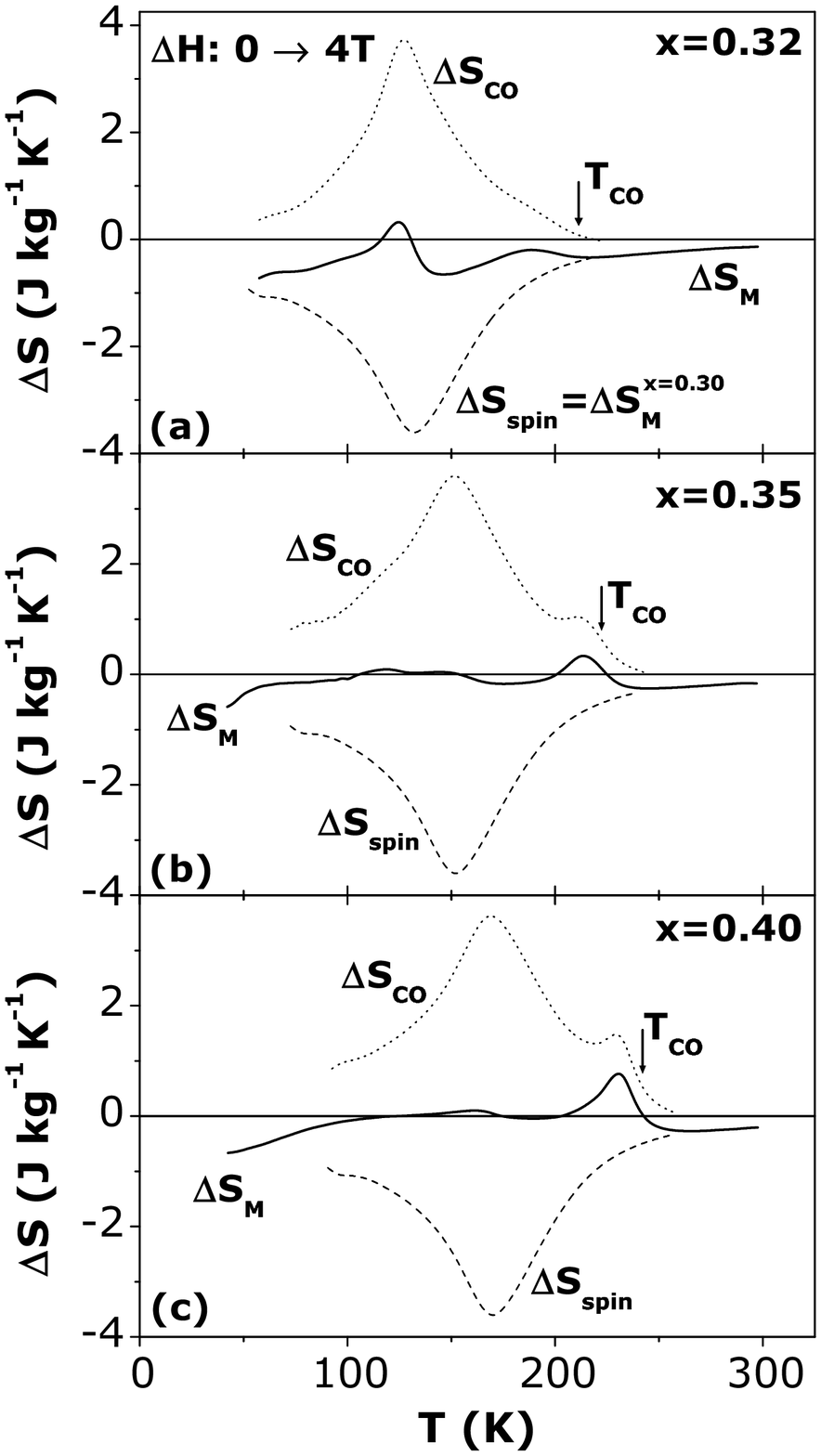}
\end{center}
\caption{Spin contribution $\Delta S_{\textnormal{\tiny{spin}}}$
and charge-ordering contribution $\Delta
S_{\textnormal{\tiny{CO}}}$ to the total magnetic entropy change
$\Delta S_{\textnormal{\tiny{M}}}$, for (a)$x=0.32$, (b)$x=0.35$,
and (c)$x=$ 0.40, and under 4 T of magnetic field change.}
\label{figure9}
\end{figure}

The behavior of the positive charge-ordering contribution, that
peaks at T$_N$, can be understood as follow. For
T$_N<$T$<$T$_{CO}$, i.e., in the paramagnetic phase, the applied
magnetic field force a rude alignment of the spins, increasing the
Mn$^{3+}$-Mn$^{4+}$ electron hopping and decreasing the
concentration of Mn$^{3+}$-Mn$^{4+}$ charge-ordered, when compared
with the zero field case. Consequently, the entropy due the CO
increase under an external applied magnetic field, allowing an
positive CO entropy change. However, for temperatures immediately
below T$_N$, the applied magnetic field favors the increasing of
the antiferromagnetic spin arrangement, comparing to the case
without field, and, consequently, the decreasing of the
Mn$^{3+}$-Mn$^{4+}$ electron mobility. Thus, the concentration of
Mn$^{3+}$-Mn$^{4+}$ charge-ordered increases, implying in the
decreasing of the entropy change due the charge-order, under an
applied magnetic field.

The magnetic entropy change for several values of magnetic field
change ($\Delta$H: 0$\rightarrow$ 1; 2; 3 and 4 T) are sketched in
figure 10, for $x=$0.32. Both contributions: charge-order
(estimated) and spin ($x=$0.30), follow the usual magnetic field
dependence.
\begin{figure}
\begin{center}
\includegraphics[width=10cm]{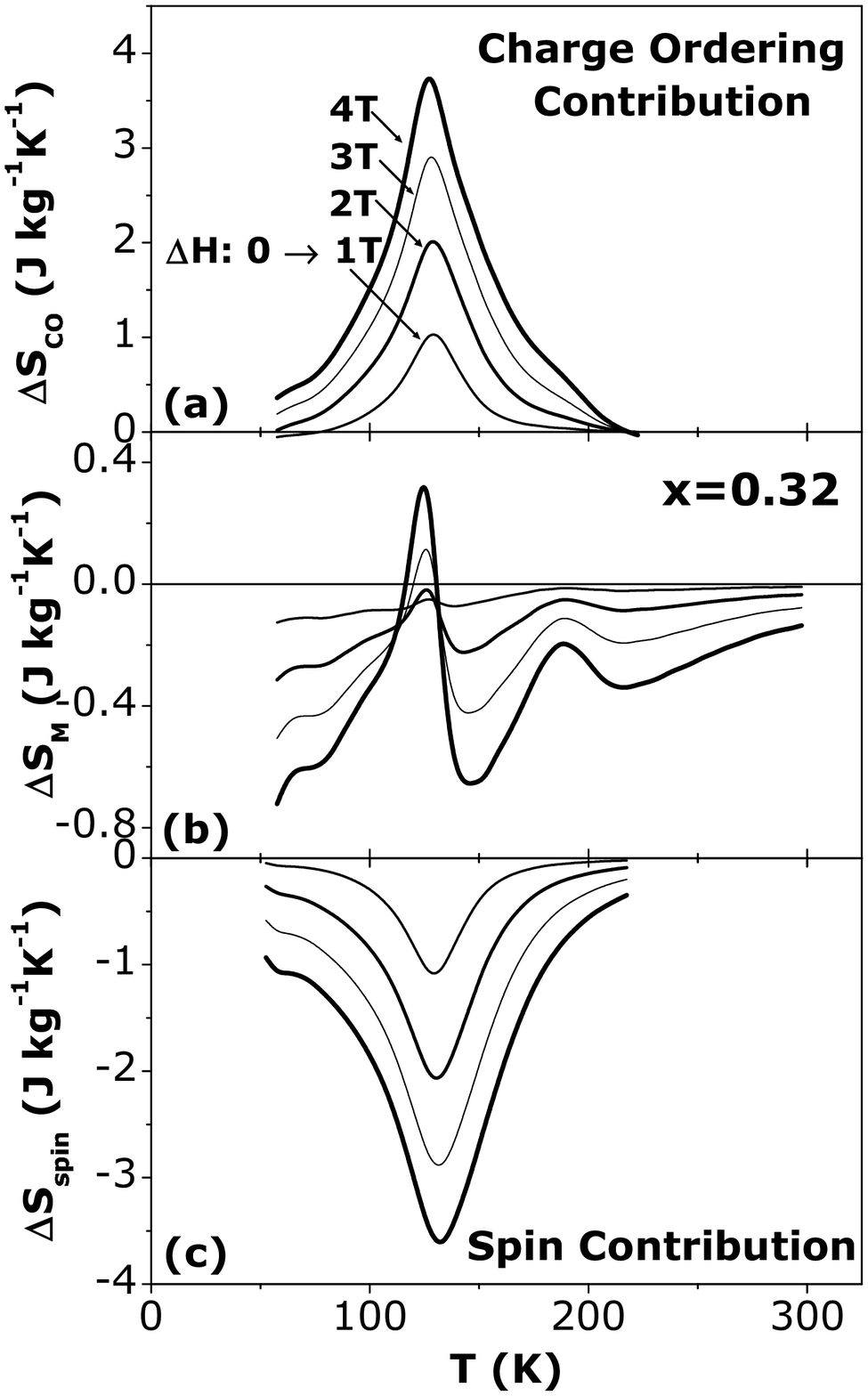}
\end{center}
\caption{Temperature dependence of the (a) charge-ordering
contribution $\Delta S_{\textnormal{\tiny{CO}}}$ and (c) spin
contribution $\Delta S_{\textnormal{\tiny{spin}}}$ to the (b)
total magnetic entropy change $\Delta S_{\textnormal{\tiny{M}}}$,
under several values of magnetic field change and $x=$0.32.}
\label{figure10}
\end{figure}

\subsection{MCE around T$^\star$}

The larger values of magnetic entropy change were obtained around
T$^\star$, where, again, the charge-ordering features found in
these manganites plays a decisive role to the MCE (see figure 1).
The value of $\Delta S_{\textnormal{\tiny{M}}}$ vanishes exactly
at T$^\star$, being highly negative (positive) for higher (lower)
temperatures. For $x=$0.32, for instance, $\Delta
S_{\textnormal{\tiny{M}}}$ reach -19.4 J kg$^{-1}$ K$^{-1}$ at 32
K, and 13.4 J kg$^{-1}$ K$^{-1}$ at 14 K. Figure 11 sketched these
features for $x=$0.32, 0.35 and 0.40, under 4 T of magnetic field
change. For the sake of clearness, table II presents the larger
values of $\Delta S_{\textnormal{\tiny{M}}}$ found in these
samples, comparing with reported values of others metals and
manganites.
\begin{figure}
\begin{center}
\includegraphics[width=10cm]{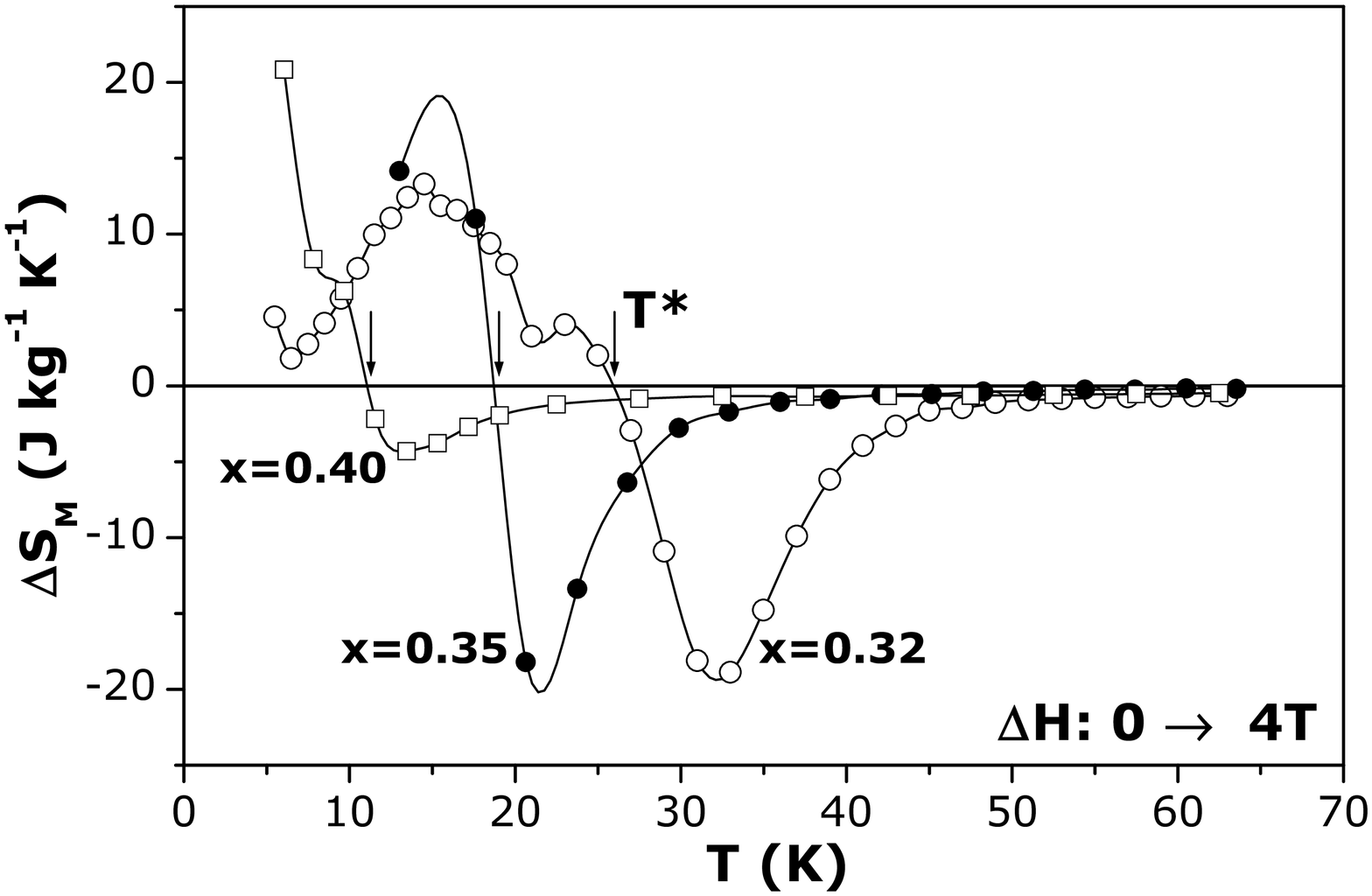}
\end{center}
\caption{Magnetic entropy change, around T$^\star$, as a
temperature function, for $x=$0.32, 0.35 and 0.40, under 4 T of
magnetic field change.} \label{figure11}
\end{figure}

\begin{table*}\label{DS}
\caption{Comparison of reported values of the maximum magnetic
entropy change of metals and manganites.}
\begin{ruledtabular}
\begin{tabular}{lcccc}
Material & $-\Delta S_{\textnormal{\tiny{M}}}$ (J kg$^{-1}$
K$^{-1}$) & $\Delta$H (kOe) & T$_C$ (K) & Ref.\\
\hline
La$_{0.67}$Ca$_{0.33}$MnO$_3$ & 6.4 & 30 & 267 & [\cite{JMMM_219_2000_183}] \\
La$_{0.60}$Y$_{0.07}$Ca$_{0.33}$MnO$_3$ & 1.5 & 30 & 230 & [\cite{APL_69_1996_3596}] \\
La$_{0.80}$Ag$_{0.20}$MnO$_3$ & 3.4 & 30 & 270 & [\cite{JMMM_222_2000_110}] \\
\textbf{Pr$_{0.60}$Ca$_{0.40}$MnO$_3$} & \textbf{4.3} & \textbf{40} & \textbf{13}\footnotemark[3] & \textbf{present work} \\
\textbf{Pr$_{0.65}$Ca$_{0.35}$MnO$_3$} & \textbf{20.1} & \textbf{40} & \textbf{21}\footnotemark[3] & \textbf{present work} \\
\textbf{Pr$_{0.68}$Ca$_{0.32}$MnO$_3$} & \textbf{19.4} & \textbf{40} & \textbf{32}\footnotemark[3] &\textbf{present work} \\
Dy & 19.5 & 65 & 174 & [\cite{JAP_77_1995_3528}] \\
Gd & 7.1 & 30 & 294 & [\cite{PRB_57_1998_3478}] \\
Gd$_{0.73}$Dy$_{0.27}$& 10 & 50 & 265 & [\cite{APL_70_1997_3299}] \\
Gd$_5$(Si$_2$Ge$_2$)\footnotemark[1]& 7 & 50 & 300 & [\cite{JAP_85_1999_5365}] \\
Gd$_5$(Si$_2$Ge$_2$)\footnotemark[2]& 14 & 20 & 276 & [\cite{JAP_85_1999_5365}] \\
\end{tabular}
\end{ruledtabular}
\footnotetext[1]{Prepared using commercial purity Gd (95-98\%
pure)} \footnotetext[2]{Prepared using high purity Gd
($\sim$99.8\% pure)} \footnotetext[3]{Around T$^\star$, instead
T$_C$}
\end{table*}

However, the positive $\Delta S_{\textnormal{\tiny{M}}}$ below
T$^\star$ can assume different values, depending on the process in
which the magnetization data were obtained. As already discussed
in section IV, the M \emph{vs.} T curves, build from several M
\emph{vs.} H curves, have different features depending if the
experimental setup performs or not a magnetic field oscillation,
around zero field, to avoid residual field on the coils for the
next M \emph{vs.} H curve (see figure 7(a)). Thus, figure 12
presents the magnetic entropy change, under 4 T of magnetic field
change and $x=$0.32, with both cases: \emph{with} and
\emph{without} the oscillating magnetic field. For the last case,
the positive $\Delta S_{\textnormal{\tiny{M}}}$ is almost
suppressed, and this feature are probably related to the metallic
state in which is the sample (for T$<$T$^\star$), since an applied
magnetic field induce a completely irreversible insulator-metal
transition for temperatures below T$^\star$, as already discussed
in section II.
\begin{figure}
\begin{center}
\includegraphics[width=10cm]{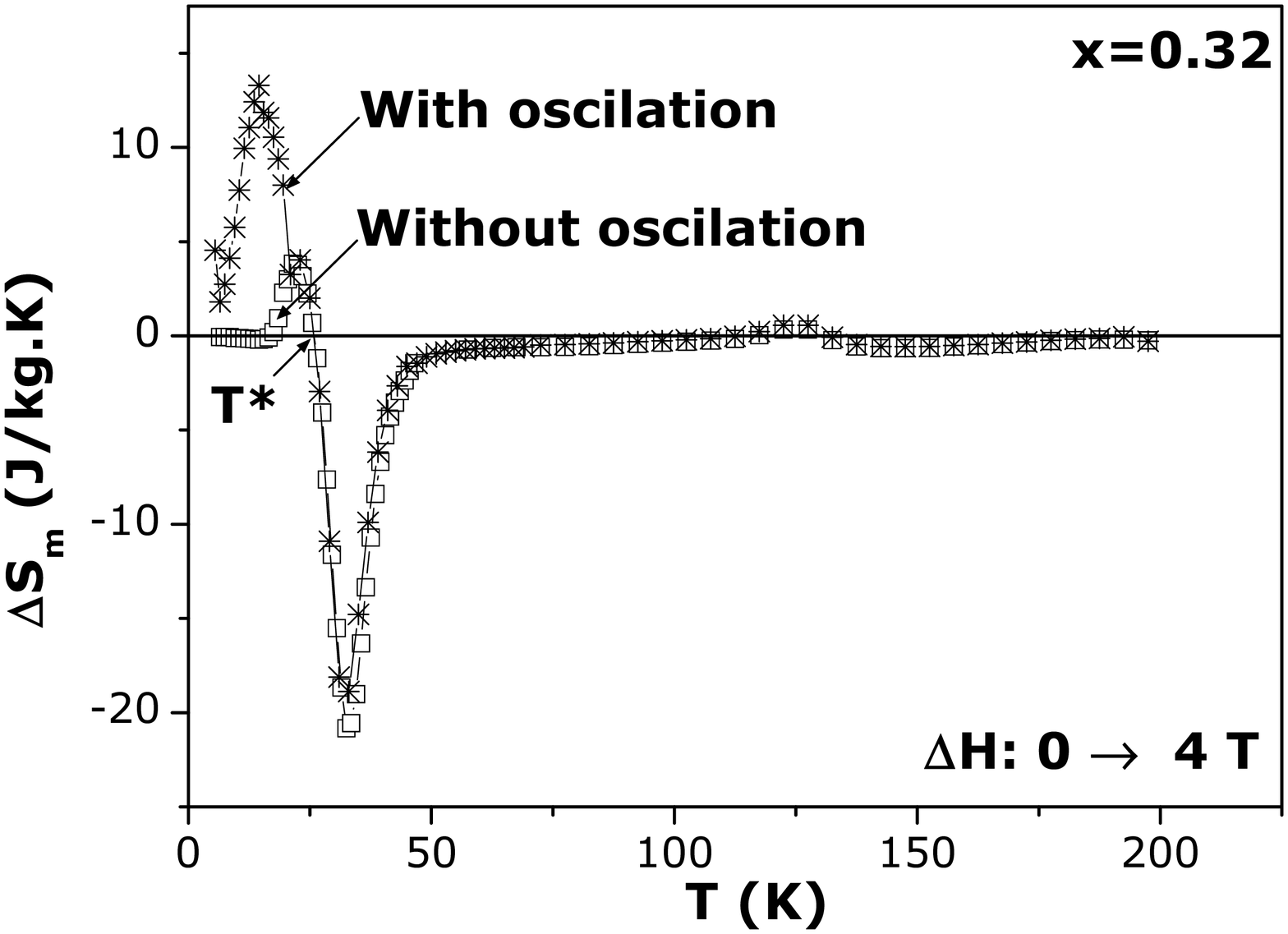}
\end{center}
\caption{Temperature dependence of the magnetic entropy change
around T$^\star$, for $x=$0.32 and under 4 T of magnetic field
change. The two cases represent \emph{with} and \emph{without}
oscillating magnetic field after each M \emph{vs.} H curve.}
\label{figure12}
\end{figure}

The temperature dependence of the magnetic entropy change has an
usual tendency with respect to several values of magnetic field
change, as presented in figure 13, for $x=$0.32 and $\Delta$
H:0$\rightarrow$1; 2; 3 and 4 T.
\begin{figure}
\begin{center}
\includegraphics[width=10cm]{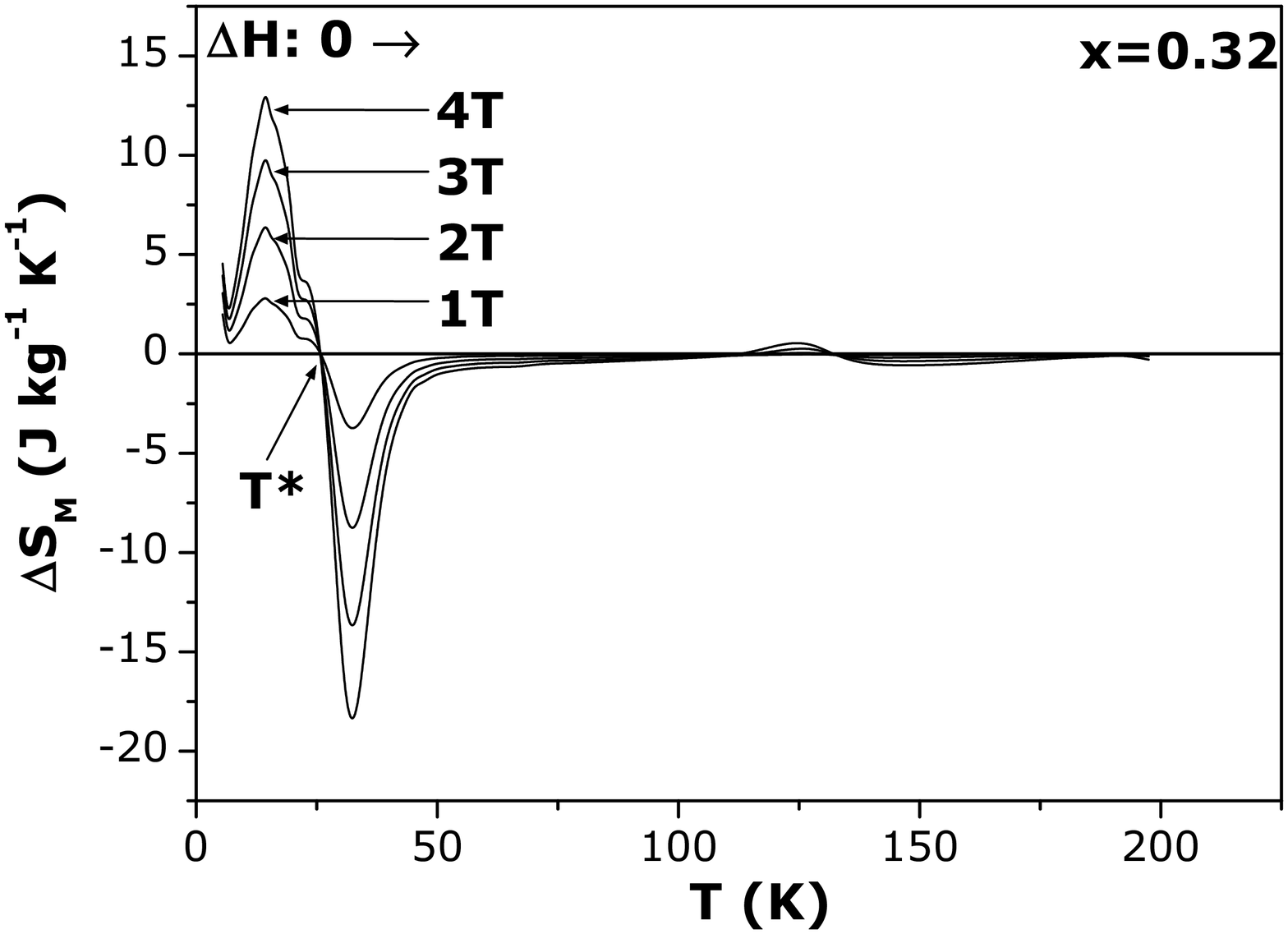}
\end{center}
\caption{Temperature dependence of the magnetic entropy change
around T$^\star$, for $x=$0.32, under several values of magnetic
field change, and for the case with oscillating magnetic field
after each M \emph{vs.} H curve.} \label{figure13}
\end{figure}

\section{Conclusion}

In the present work, we found an anomalous magnetic entropy change
for $x>$0.30 (concentrations exhibiting charge-ordering
phenomenon).The results could be explained considering a spin and
charge-ordering contributions to the total magnetic entropy
change. Moreover, we found an extremely large value for the
entropy variation, that occurs at a characteristic temperature
T$^\star$. Other manganites showing the characteristic temperature
T$^\star$ (see ref. \cite{JMMM_200_1999_1}), can also present
large magnetic entropy change, and, consequently, a great
potential to be employed in various thermal devices.

In some previous publications
\cite{EL_58_2002_42,{PRB_66_2002_134417},{langevin}} we pointed
out that the unusual properties of manganites result from their
magnetic non-extensivity, in the sense of Tsallis statistics
\cite{PA_261_1998_534}. Such an approach is being applied to the
analysis of the present results and will be published elsewhere.

\begin{acknowledgments}
The authors thanks FAPERJ/Brasil, FCT/Portugal (contract
POCTI/CTM/35462/99) and ICCTI/CAPES (Portugal-Brasil bilateral
cooperation), for financial support.
\end{acknowledgments}

\end{document}